# Synthesis of europium-based crystals containing As or P by a flux method: attempts to grow EuAgP single crystals

Karolina Podgórska[1,*], Damian Rybicki[1], Lan Maria Tran[2], Wojciech Tabiś[1], Łukasz Gondek[1], Michał Babij[2,**]

[1]AGH University of Krakow, Faculty of Physics and Applied Computer Science, al. A. Mickiewicza 30, 30-059 Kraków, Poland

[2]Institute of Low Temperature and Structure Research, Polish Academy of Sciences, Okólna 2, 50-422 Wrocław, Poland

## Abstract

Europium-based materials are highly attractive due to their diverse range of physical properties. In these studies, we aimed to synthesize single crystals of the potentially topological semimetallic compound EuAgP, which up to this day has only been obtained in polycrystalline form. The flux method was employed for the syntheses, using fluxes such as: Bi, Sn, Pb, and In, in their various ratios. The purpose of using Bi flux was to try synthesizing an analog of EuAgAs single crystals, by fully substituting arsenic with phosphorus. The obtained crystals were characterized by x-ray diffraction and scanning electron microscopy. Despite many unsuccessful attempts to synthesize EuAgP single crystals, the study provides valuable insights into how different fluxes and their ratios influence the final synthesis product. It also underscores the complexity of designing analogs between arsenides and phosphides.

## 1 Introduction

There are several methods of crystal growth: solid state technique, chemical vapor transport, and solution methods. Often the first attempt to synthesis of single crystals is utilization of the so-called solid state technique, where the crystals are grown directly from a mixture of solids. For a mixture of congruently melting substrate the reaction can easily result in mm to cm single crystal. For example nearly congruently melting Fe-Se-Te system results in mm sized crystals [1] or really congruently melting Ga-As system, which are cm sized [2]. In many other cases the disadvantage of this technique is that the obtained crystals may be twinned and have grain boundary problems. In contrast, the chemical vapor transport method allows obtaining good quality crystals. This method involves the transport of material using a transport agent, typically iodide or chloride, in vapor form from a hot zone to a slightly cooler zone in a sealed vessel where crystallization occurs. However, this method presents challenges, such as: identifying the right transport agent, determining the right temperature gradient, and the extended time required for the crystal to grow to the desired size [3]. Solution growth methods can be categorized into two main types: those that rely on slow and controlled evaporation of the solvent from the solution, and those that induce crystallization by altering the temperature to change the solubility of individual substances within the solution, known as the solvothermal method. A specific variant of the solvothermal method is the flux method of synthesis [4], which is widely used to obtain inorganic compounds in a form of single crystals. The flux methods can be divided by the type of flux used, e.g., salt (salt flux), metal (metal flux) or excess of one of the synthesized compound's precursors (self-flux). It should be noted that flux is not necessarily a solvent and does not have to dissolve or react with substrates and synthesis products. Instead, it acts as an environment that allows crystals to grow during slow heating or cooling.

*Corresponding author
**Corresponding author
E-mail addresses: klpodgor@agh.edu.pl (K. Podgórska), m.babij@intibs.pl (M. Babij)

Compounds containing Eu show a wide range of distinct physical properties, i.e., they can: behave like insulators, metals or semiconductors, sometimes showing colossal magnetoresistance [5]; form heavy fermion systems with Kondo effect [6]; have optical properties such as absorption-emission [7]; be high temperature superconductors with unusual behavior, like Chevrel phases [8] with Jaccarino-Peter compensation effect or doped $EuFe_2As_2$ iron based superconductors [9–11]; or even be topological materials [12]. Many of these interesting properties are a consequence of the two possible valance states of Eu: the "magnetic" $Eu^{2+}$ with $4f^7$ configuration ($L = 0$, $S = J = 7/2$, $\mu_{eff} = 7.94$ $\mu_B$) and the "non-magnetic" $Eu^{3+}$ with $4f^6$ configuration (with $L = S = 3$, $J = 0$, $\mu_{eff} = 0$ $\mu_B$). While compounds with $Eu^{3+}$ and $Eu^{2+}$ can be optically active [7], only the $Eu^{2+}$ compounds are capable of long-range magnetic order [11,13,14], which can be further influenced by chemical doping [15,16], or external pressure [14]. In some materials an intermediate valance state is also observed [17,18], and the modification of valance state can occur with temperature change [19] or applied external pressure [20–22]. The possibility of magnetic order in Eu-based materials compared to non-magnetic counterparts (such as Ca- or Sr- based) gives an opportunity to investigate the influence of magnetism on the structure and properties of these materials [23].

Single crystals are often the preferred form of material for study, since they allow to study the anisotropy of the compound, e.g., magnetic properties can differ significantly (even in orders of magnitude) depending on the orientation of the single crystal with respect to the magnetic field. Single crystals are also typically of higher quality than polycrystalline samples. Unfortunately, obtaining good quality crystals can be a difficult task. Even if one synthesizes crystals using the flux method, the properties of grown crystals may depend on the type of the flux. This was the case for example in Eu(Fe$_{1-x}$Co$_x$)$_2$As$_2$, while the Néel temperature was nearly unchanged, the superconducting critical temperature was twice as low for samples grown from Sn flux compared to the self-flux crystals with the nearly the same Co-doping ($x$) [15,24]. Perversely, it was reported that for self-flux syntheses there was a "lack of superconductivity in the phase diagram of single-crystalline Eu(Fe$_{1-x}$Co$_x$)$_2$As$_2$" [25], which was true as long as the samples did not undergo normalization process. Moreover, crystals obtained from the same batch can differ [26]. Therefore, it is not surprising, that even if the same flux is used and synthesis conditions are very similar and high-quality single crystals are obtained, properties of the grown material can be somewhat different. Such is the case for EuZn$_2$As$_2$ or EuZn$_2$P$_2$. While their respective Néel temperatures reported by different groups were similar, their detailed magnetic characteristics varied leading to different conclusions on the magnetic structure of these compounds [12–14,27–29].

The aim of this study was to synthesize single crystals of EuAgP, a potential topological semimetallic compound, previously only obtained in polycrystalline form [30,31]. For this purpose, a flux method was chosen. Various types of fluxes, including Bi, Sn, Pb, and In, as well as different ratios of these fluxes were tested. Additionally, we provide detailed information on the entire synthesis process and equipment, which can be helpful for researchers interested in preparing single crystals.

## 2  Methodology

The substances and equipment detailed below were used for the synthesis and characterization of the obtained crystals described in the subsequent sections. Weighing of individual substances was carried out in a glovebox under a protective argon atmosphere.

During different attempts the following elements were used as reagents: Eu (Onyxmet, dendritic, 99.99%, in mineral oil), Ag (Specpure spectrographically standarised), P (Sigma Aldrich, 99.99%, red phosphorus), As (Alfa Aesar, 99.999%), and as fluxes: Bi (KOCH-LIGHT, 99.999%), Sn (purity 99.999%), Pb (Onyxmet, 99.999%), In (purity 99.999%). Weighed substances were placed in alumina crucibles placed in quartz ampules. The prepared ampules were evacuated using the turbomolecular pump (Agilent Technologies, TPS-compact), and then flame sealed. The sealed ampules were heated in Nabertherm (Mod. N 20/14) chamber furnace. To separate the crystals from the flux after the synthesis was completed, a modified centrifuge (MPW Med. Instruments, model M-diagnostic) was used.

The obtained crystals were characterized using x-ray powder diffraction (XRD; Panalytical Empyrean diffractometer) and x-ray microanalysis (EDX; FEI Nova Nano 230 scanning electron microscope with EDX option).

# 3 Flux method

## 3.1 Flux selection

The success of the flux synthesis method depends on numerous factors. One of them is selection of the correct flux, which acts as a liquid medium that enables solid-phase reactions to take place. As a metallic flux elements such as indium, lead, bismuth, tin and antimony are mainly used. An advantage of these metals is their relatively low melting point (below 500 ℃), which allows a lot of elements to dissolve in the flux more easily.

Analyzing the phase diagrams of individual elements and flux can help in determination whether a particular flux is suitable for a specific synthesis. During crystal growth, it is important to follow the lines of liquidus. From the phase diagrams, one can also estimate how well a substance dissolves in the flux. If a substance dissolves too readily, it may completely dissolve without crystallizing. In such cases, adding an excess of the substance can be helpful. Conversely, if a substance dissolves poorly, additional flux may be needed to promote dissolution and crystallization. This consideration is particularly significant when creating an analog compound by replacing one element with another, such as substituting arsenic with phosphorus, or changing the type of doping in the parent compound. For example, when preparing the synthesis of $EuFe_2As_2$ doped with cobalt the used elements are in nearly stoichiometric ratios [32], in contrast when doping $EuFe_2As_2$ with nickel it is necessary to add ten times greater amount of nickel to obtain crystals with the desired stoichiometry [16]. Each element may dissolve differently in a specific flux, so the ratios used for synthesis should be carefully adjusted accordingly. Typically, for flux methods, around 20-30 g of flux is used, while the mass of the sample is usually in the range 1-1.5 g.

It is important to remember that even when crystals are successfully obtained, verifying their composition is crucial, as the flux may readily build into the crystal structure. In more complex scenarios, flux can also get trapped in defects, interstitial sites, or between the planes of twinned crystals, leading to altered physical properties. For example, in the case of $BaFe_2As_2$, using Sn flux instead of FeAs (self-flux) for synthesis, results in lowering of the spin density wave ordering temperature [33,34].

## 3.2 Crucibles

Another key factor in successful synthesis is selecting the appropriate crucible material. Crucibles come in various shapes and are made from different materials. One of the most widely used crucible materials for non-organic synthesis is alumina. However, generally for this purpose the following materials can be used: resistive metals (platinum, tantalum), oxides (quartz, alumina, zirconia), or other chemical resistive materials (such as BN). Syntheses may also be carried out directly in airtight steel container or quartz ampules (often double quartz ampules) if they need to be carried out in a non-air atmosphere. While quartz can be susceptible to corrosion, synthesis can still be performed if one uses a double quartz ampule. This approach is possible as long as the corrosion rate is slow enough to prevent the outer ampule from corroding. The inner ampule acts as diffusion barrier and the role of the outer ampule is to protect the synthesis from external conditions, such as exposure to air. In the case where the mixture is highly reactive, corrodes with almost all available crucible materials and leads to crystal growth with impurities or shift in their composition, one solution is to develop a crucible from the corrosion product itself [36]. To conclude, it is recommended to select a crucible material that does not react with the reaction mixture. A more detailed discussion on the topic of crucibles is also presented in Ref. [3,35].

The preparation of a reaction mixture involving air-unstable or toxic reagents has to be carried out in a glovebox with a protective atmosphere to ensure safety and prevent contamination. The stoichiometric ratios of the elements of the compound to be synthesized and the flux should be weighed and placed in a crucible in a quartz tube (see Figure 1(a)). The quartz ampule should then be evacuated and flame sealed. The selection of quartz as a material for ampules is not random. Diffusion quartz can withstand temperatures up to 1200 °C and remains mainly non-reactive. Additionally, its heat treatment allows for the straightforward closing of the ampules containing the prepared ingredients. Moreover, quartz can withstand partial pressures of about 2 atm in larger ampules (see Figure 1(b,c)) sealed under vacuum, and up to 8 atm in small ampules (see Figure 1(d)). The safe pressure limit is approximately 2 atm at 400 °C and 700 °C for phosphorus and arsenic, respectively, making quartz well-suited for synthesizing compounds containing these elements. Due to this pressure limit, it is essential to consider the amount of material placed inside the crucible and sealed in a quartz ampule. For example, for some arsenides, preparing 7 g of material in a single ampule can lead to an explosion of the ampule during heat treatment procedure, while reducing the ratio to about 2 g is sufficient to make the synthesis successful. Another crucial aspect to consider is whether the ampule is under vacuum or contains residual gas, as this can greatly influence the resulting product of synthesis. Variations in these conditions can lead to differences in crystal growth even when other factors are supposedly identical.

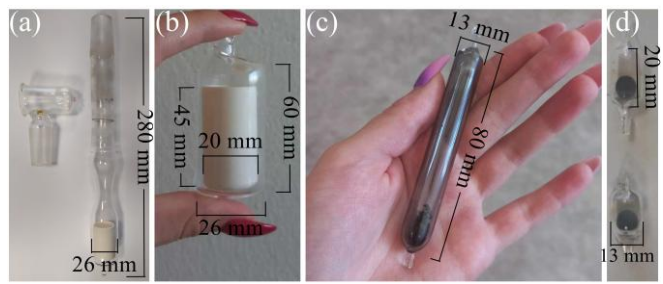

*Figure 1 (a) Empty quartz tube prepared for flame sealing and (b-d) examples of flame sealed ampule designs in various sizes that effectively minimize the risk of explosions during synthesis: (b) large ampule with alumina crucible before heat treatment, (c) large ampule after heat treatment with synthesis product, (d) small ampules before heat treatment with substrates pressed into pellets. In this work, the syntheses were conducted using the quartz tubes and ampules presented in sub-figures (a) and (b). The approximate dimensions of the quartz tube and individual ampules are given in the figures, with a quartz thickness of 1 -2 mm.*

## 3.3 Heating profile

Another factor that plays a significant role in the successful synthesis is the choice of appropriate heating profile. First, the ampule must be heated above the liquefy temperature of the substances used, usually this step takes 10-15 h. Then, to ensure that there is homogeneous liquid in the crucible, the ampule is kept at the maximum temperature for several hours (~10 h). This is followed by a much longer (160-270 h) cooling process during which the actual crystal growth occurs. Sometimes, to obtain larger crystals a special procedure is undertaken. At first, a slow heating rate is set to form many crystals during the heating process. Then, the temperature is raised until most of the crystals melt and only a few remain. The remaining ones will act as seed crystals and will continue to grow during the following slow cooling process.

When elements such as phosphorus or arsenic are used in the synthesis, it is important to remember that an incorrect choice of heating profile can result in an explosion of the ampule in the furnace. For such syntheses, it can be helpful to hold the furnace temperature at 400 °C for phosphorus and at 500 °C for arsenic for several hours during the heating process.

## 3.4 Furnace

For the heating process, the sealed quartz ampules are placed in the furnace. There are many types of furnaces, and some are adapted to very specific syntheses. Some of the more universal and popular are two types of furnaces: tube and chamber one.

In the case of the first type the heating zone is quite small and the temperature gradient nearly always occurs. Therefore, this type of furnace is very suitable for short ampules. Also tube furnace allows to conduct synthesis only for one sample (ampule) at a time. However, for obtaining good quality large crystals it could be more advantageous since the small temperature gradient allows to control direction of crystal growth in the crucible.

This is the reason why an extremely popular variation of tube furnaces are multi heat zone furnaces, which have two or more separate heating elements (zones) close to each other. This allows to provide huge gradient of temperature in one ampule (reacting vessel). This type of furnace is dedicated for chemical transport growth of crystals.

The second type is a chamber furnace, in which inner space is suitable for multiple samples (ampules) and their length (height) is less important. Additionally, there is nearly no temperature gradient in the chamber space, therefore the whole ampule has the same temperature. However, it needs to be pointed out that sometimes large chambers might have hotter/colder regions (near heat element or door). Nevertheless, the entire ampule will have the same temperature.

## 3.5 Crystals extraction

After synthesis, the crystals need to be separated from the flux. One of such methods is centrifugation of the still-hot ampules removed directly from the furnace, while the flux is still in a liquid state [3]. However, for syntheses carried out for this research a somewhat different approach was used.

In contrast to the method in Ref. [3] the ampules were first cooled down to room temperature according to the set heating (cooling) profile. By opening the cooled ampules, the crucibles containing crystals trapped in solidified flux were retrieved. The crucibles were then placed inside new quartz ampules. As a strainer, on which the crystals stop during centrifugation, an appropriate amount of quartz wool was deposited inside, followed by a small amount of quartz cullet on top (an alumina strainer can be used instead).

The prepared ampules (see Figure 2(a)) were first evacuated, flame-sealed, and then heated for approximately half an hour at a temperature at which the flux is already in liquid state. The still-hot ampules were removed from the furnace and placed upside down in the centrifuge. A centrifuge for large-scale biological samples with certain modifications (see Figure 2(b,c)) was used. In this type of centrifuge, it is possible to centrifuge the ampules heated up to 800 °C. As the ampule spins the bigger crystals are collected on the "strainer", and due to the centrifugal force, the flux is separated. After centrifugation, the ampules with separated fractions (crystals and flux) were cooled and then opened.

In contrast to the method described in Ref. [3], our approach involves cooling the ampules with the synthesis products to room temperature before separation. This method eliminates the need for two crucibles, making it more cost-effective. However, it requires the preparation of a new setup specifically for centrifugation.

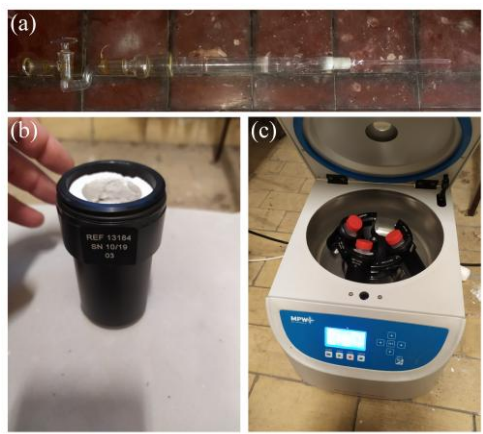

*Figure 2 (a) Ampule with crucible, quartz wool, and quartz cullet. (b) Centrifugation container in which hot ampule is placed. (c) Modified centrifuge for hot ampule centrifugation.*

Despite using this procedure, some amount of flux may still be present on the surface of the obtained crystals. As a result, during measurements such as magnetic susceptibility a residual signal from the flux can be observed. For example, a diamagnetic signal resulting from a superconducting transition of tin may be detected below 3.7 K. However, there are effective methods to address the residual flux on the surface of crystals. It can be removed chemically

(using acetic or hydrochloric acid), or mechanically (by grinding with sandpaper or sanding paste). Nevertheless, the least invasive method of removing the flux is the amalgam method. This method was employed to clean the crystals obtained in this work. The amalgam method involves immersing the crystals in metals that will most likely form an amalgam with the flux. Usually, metals with approximately room temperature melting point, such as gallium (which is already a liquid at 30 °C) are used. The amalgam on crystals' surface is then removed by centrifugation. Note that the amalgams are centrifuged, not evaporated, as the flux will remain on the sample after evaporation.

# 4 Route to synthesis of EuAgP

In this study, attempts were made to synthesize EuAgP, that, according to the literature, has previously only been obtained in a polycrystalline form [30,31]. Notably, a similar compound, EuAgAs, can be obtained from Bi flux [37,38]. These two compounds differ by only one element and crystalize in the same crystal structure (space group *P*6$_3$/*mmc*). Arsenic is the chemical analogue of phosphorus, i.e., both elements have the same number of valence electrons, almost the same electronegativity and orbital configuration. These similarities between phosphorus and arsenic, lead us to believe that single crystals of EuAgP could be also obtained from Bi flux. However, it is important to note that As and P have different ionic radii, meaning they vary in size. As a result, achieving an exact synthesis analogue for compounds differing only by these two elements may not always be possible.

## 4.1 Bi flux

As discussed in Section 1, obtaining high-quality crystals with the desired composition is challenging due to the numerous factors that influence the final synthesis product. In our efforts to synthesize crystals of EuAgP, we initially replicated the synthesis of EuAgAs in our experimental setup to determine if we could achieve the same results. The required elements: Eu, Ag, As, and Bi were weighed in a ratio of 1:1:1:9 in an Ar glovebox and placed in an alumina crucible. The amount of reactants placed inside the crucible was limited by its capacity. In this synthesis attempt, 0.695 g Eu, 0.490 g Ag, 0.340 g As, and 8.549 g Bi were used to obtain 1.5 g of the final product (assuming 100% yield). The quartz tube with crucible inside was then evacuated and flame sealed. The prepared quartz ampule was heated in the Nabertherm chamber furnace for 15 h to a temperature of 1100 °C and kept at this temperature for 24 h. Subsequently, cooling process was started, and was carried out for 250 h to a temperature equal to 700 °C (see Figure 3). Finally, the ampule was cooled to room temperature, opened and crystals were extracted in accordance with the process described in Section 3.5. The synthesis resulted in black plate-shaped crystals, see Figure 4(a), their stoichiometry was confirmed by energy dispersive x-ray spectroscopy (EDX) measurement (see Figure 5(a)). Example crystal photo taken using scanning electron microscope (SEM) is shown in Figure 6(a) and the composition details are summarized in Table 1. A map of the elemental distribution for an exemplary crystal was also made and is presented in Figure 7(a). The distribution of individual elements is uniform. The observable black areas on the sample are flux remains on the surface (cf. circled areas in Figure 7(a)).

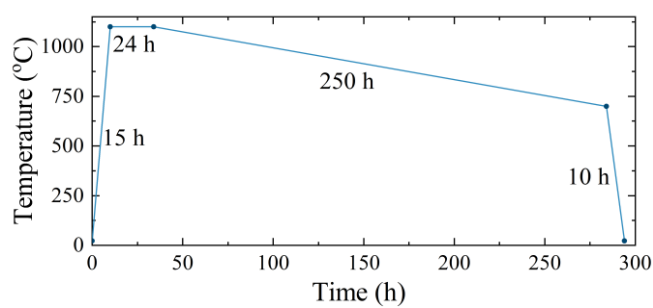

Figure 3. The temperature profile used for the syntheses described in this work.

For the mentioned crystals, x-ray diffraction (XRD) measurements were also performed. XRD data with Rietveld refinement [39] for EuAgAs are shown in Figure 8(a). XRD pattern collected for EuAgAs was indexed and refined using the hexagonal $P6_3/mmc$ space group. At 300 K the refined lattice parameters are $a$ = 4.5122(1) Å and $c$ = 8.1022(2) Å, which is consistent with the literature data [31].

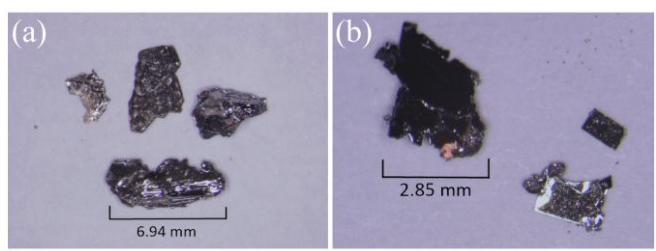

Figure 4. Crystals of (a) EuAgAs obtained from Bi flux and (b) EuSnP (obtained in the attempt to synthesize EuAgP) from Sn flux.

Following the successful synthesis of the EuAgAs, we proceeded with an attempt to synthesize EuAgP using Bi flux. The metal pieces of Eu, Ag, P, Bi were weighed in the same ratio (1:1:1:9) as for the EuAgAs, and all synthesis steps were performed analogously. In this attempt, 0.510 g of Eu, 0.359 g of Ag, 0.103 g of P, and 6.269 g of Bi were used. However, despite these efforts, no crystals of EuAgP were obtained.

In another attempt, to address the possibility that the used amount of flux was insufficient, the ratio of the flux was increased from 9 to 40 (the flux weight was increased to 28.726 g). Despite this change, no crystals of EuAgP were obtained.

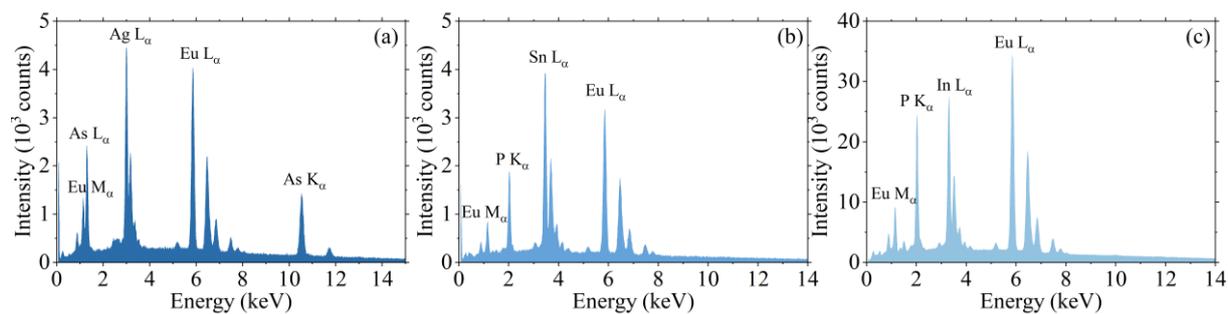

Figure 5. EDX spectra of single crystals of: (a) EuAgAs grown from Bi flux, (b) EuSnP grown from Sn flux, (c) $Eu_3In_2P_4$ grown from In flux.

Table 1. Estimated composition of synthesized single crystals based on EDX measurements on several samples.

| Sample | At% | | | | | |
|---|---|---|---|---|---|---|
| | Eu | Ag | As | P | Sn | In |
| EuAgAs | 33.93 | 32.69 | 33.39 | - | - | - |
| EuSnP | 33.61 | - | - | 32.70 | 33.69 | - |
| $Eu_3In_2P_4$ | 34.43 | - | - | 42.16 | - | 23.41 |

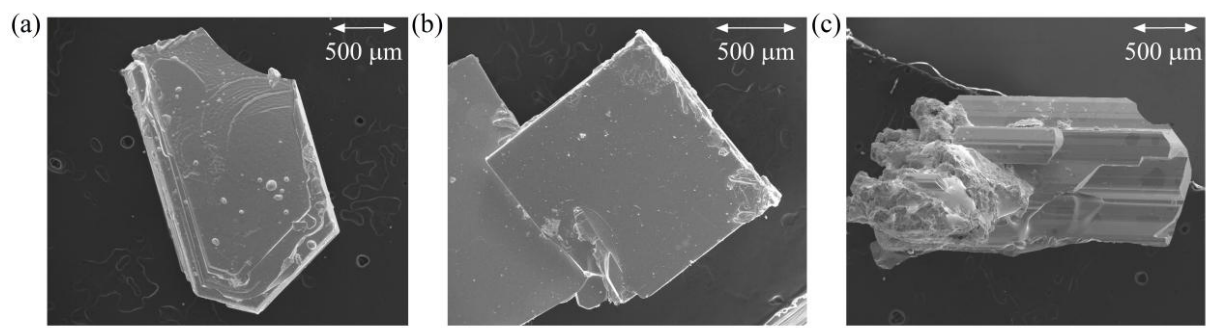

Figure 6. SEM images of single crystals of: (a) EuAgAs grown from Bi flux, (b) EuSnP grown from Sn flux, (c) $Eu_3In_2P_4$ grown from In flux.

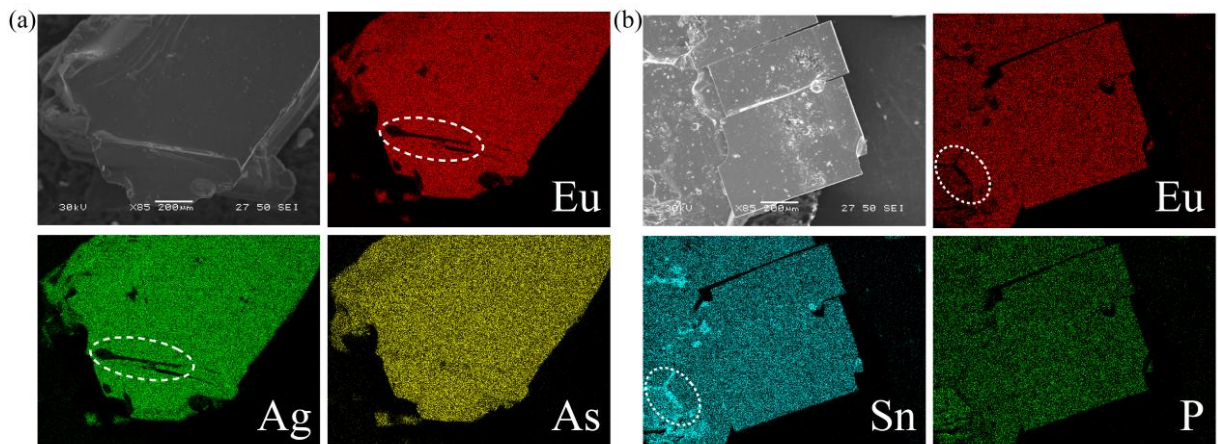

*Figure 7 Maps of the elemental distribution for example single crystals of (a) EuAgAs and (b) EuSnP. (a) Remnants of flux on the surface of: (a) EuAgAs, circled with dashed line; and (b) EuSnP, circled with dotted line.*

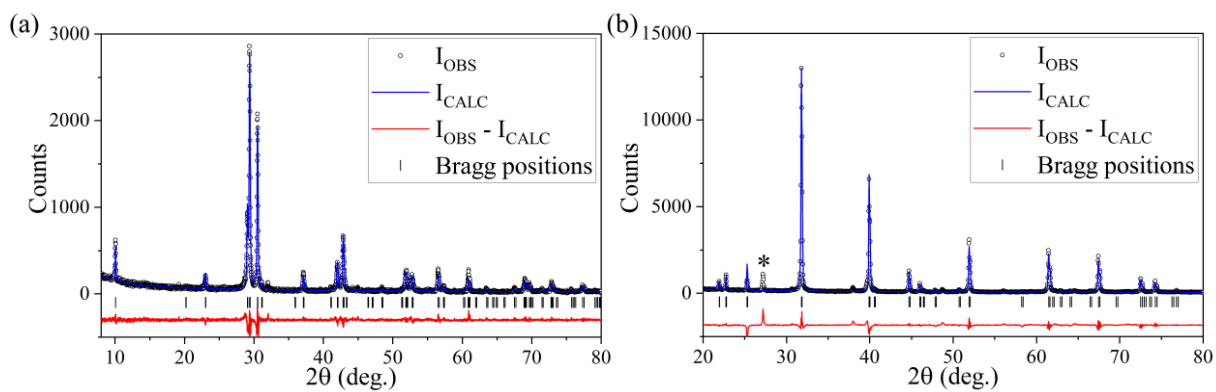

*Figure 8 X-ray diffraction data (open points) for (a) EuAgAs and (b) EuSnP with Rietveld refinement (blue solid line). Bragg positions are marked by bars and the red solid line corresponds to difference between x-ray diffraction pattern ($I_{OBS}$) and Rietveld refinement ($I_{CALC}$). Reflex at 23° of 2θ (marked as *) originates from Sn flux.*

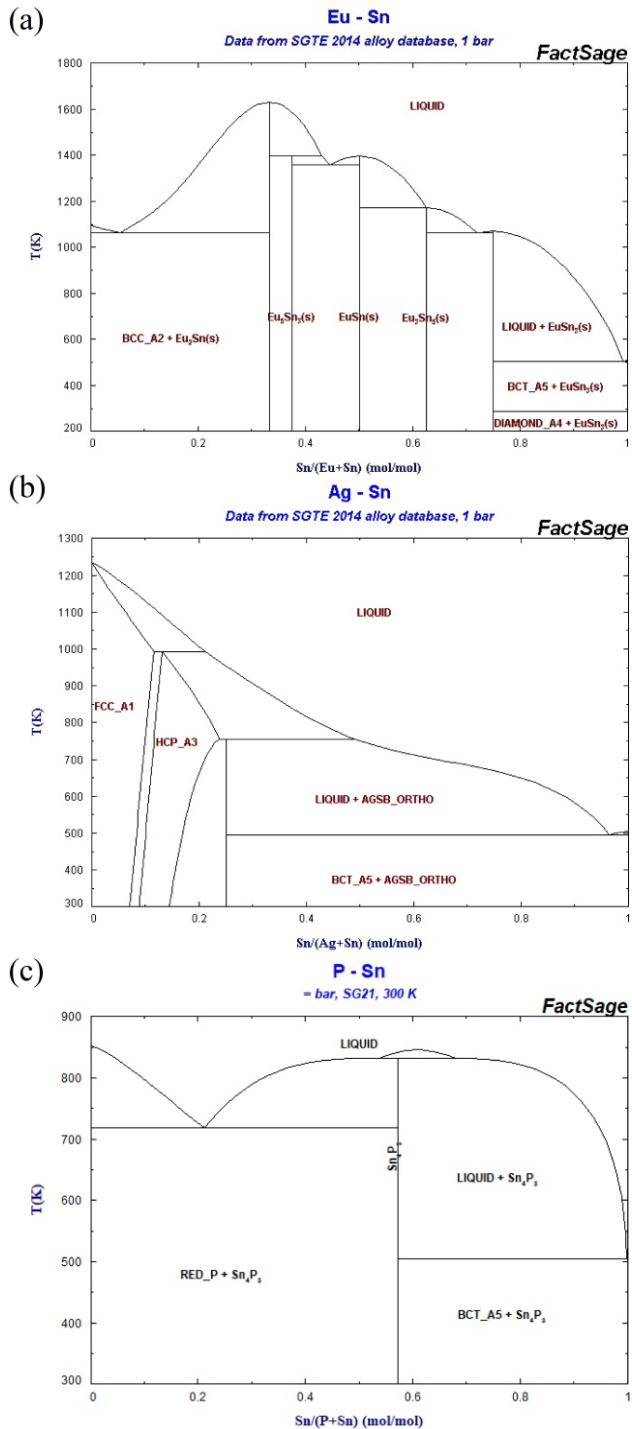

*Figure 9. Binary phase diagrams of: (a) Eu-Sn, (b) Ag-Sn, (c) P-Sn* [40]

## 4.2 Sn flux

Following the unsuccessful attempts with Bi flux, other types of metallic fluxes were investigated. The next synthesis was performed using Sn flux. However, there was a probability that Sn would enter the compound, as it is known to substitute for both silver and phosphorus in certain compounds [41,42]. During the cooling process, it was essential to stay above the liquidus line to prevent the formation of binary compounds. Analysis of the phase diagrams for

Eu-Sn, Ag-Sn, and P-Sn (cf. Figure 9) indicated that a tin-rich region would be optimal for crystal growth. Therefore, it was decided to use 30 times the amount of the Sn flux to remain within the desired area of the phase diagram. For this process, 1:1:1:30 of Eu:Ag:As:Sn ratio was used. Specifically, this involved using 0.684 g of Eu, 0.486 g of Ag, 0.139 g of P, and 16.090 g of Sn. Care was taken to ensure that the total amount of flux and reactants were within the capacity of the crucible. In this and the following tested fluxes the same synthesis conditions were applied as in the case of Bi flux, i.e., alumina crucibles in vacuum flame sealed quartz ampules were used and the synthesis process was carried out in chamber furnace with the same temperature profile (see Figure 3) as for the Bi flux, followed by crystal separation (the same as described in Section 3.5).

Upon completion the synthesis and subsequent centrifugation of the ampule, black platelet-shaped crystals were obtained (see Figure 4(b)), similarly as in the case of EuAgAs. However, EDX measurements showed that instead of EuAgP, crystals of the EuSnP were obtained (see Figure 5(b)). Apparently, Sn incorporated into the crystal structure, as initially suspected. An SEM image of an example crystal is shown in Figure 6(b) and details of the composition can be found in Table 1. A distribution map of individual elements obtained from SEM is shown in Figure 7(b). It is notable that in certain areas (circled with dotted line in Figure 7(b)), enriched by Sn, a lower concentration of Eu and P in observed, which is due to the remnants of the Sn flux on the surface of the crystal.

EuSnP crystallizes in the tetragonal P4/*nmm* space group. At the temperature of 300 K the unit cell parameters according to the x-ray diffraction result with Rietveld refinement (Figure 8(b)) are as follows: $a = 4.2979(1)$ Å and $c = 8.7848(2)$ Å. Non-indexed, small reflex (marked with * in Figure 8(b)) at 23° of $2\theta$ originates from metallic Sn flux. The obtained results are consistent with those reported in the literature, confirming the existence of EuSnP crystals synthesized by the flux method using Sn flux [43].

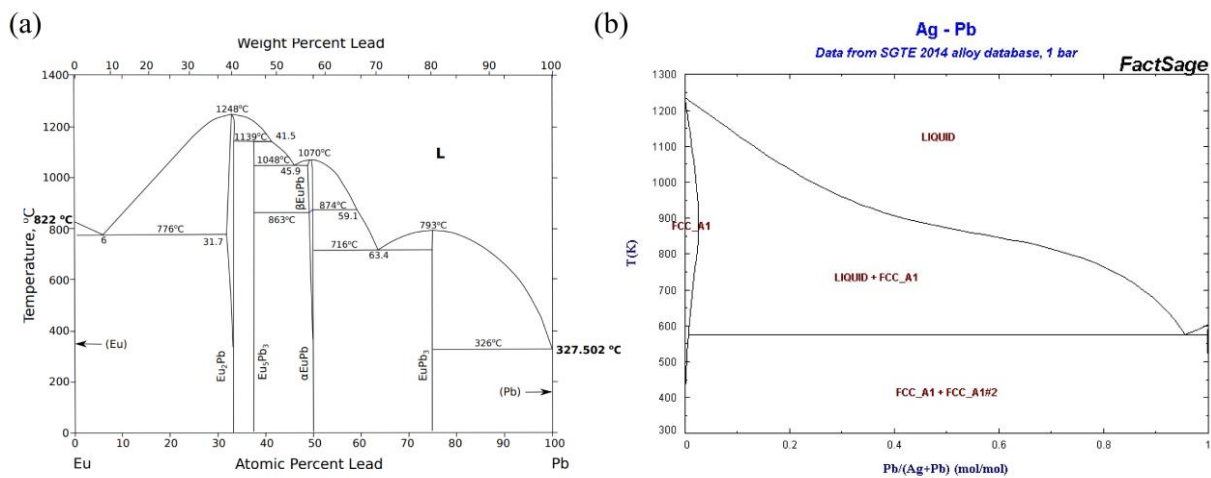

*Figure 10. Binary phase diagrams of: (a) Eu-Pb (after* [44]*), (b) Ag-Pb* [40]

## 4.3 Pb flux

The next flux tested was lead. In contrast to Sn, there are no known ternary compounds of Eu and Ag containing Pb. Unfortunately, there is no Pb-P phase diagram available in the literature. Thus, based only on two available binary phase diagrams: Eu-Pb and Eu-Ag (cf. Figure 10), the flux (Pb) rich region was chosen as the optimal area for crystal growth. Using the same procedure as for previous attempts the synthesis with 1:1:1:40 of Eu:Ag:P:Pb was carried out. In this attempt we used 0.522 g of Eu, 0.371 g of Ag, 0.106 g of P, and 28.481 g of Pb. However, no crystals were obtained from this attempt either.

## 4.4 In flux

The last flux tested was indium. Given the possibility that In could be incorporated in the crystal structure, due to the existence of many indium ternary compounds such as: EuAgIn, $Eu_3Ag_2In_9$, $EuAg_4In_8$, $Eu_3InP_3$, and $Eu_3In_2P_4$ [45], synthesis with In flux was carried out. By analyzing the binary phase diagrams (see Figure 11), it was assumed that using 1:1:1:20 Eu:As:P:In ratio would place the system in the optimal area for crystal growth. In this case, the following masses of each substance were used: 0.661 g of Eu, 0.469 g of Ag, 0.135 g of P, and 9.984 g of In. As a result, some single crystals were indeed obtained. However, the crystals contained In, as anticipated (see Figure 5(c)). Based on the EDX measurements, the stoichiometry of the obtained material is most likely $Eu_3In_2P_4$ (see Table 1). SEM image of crystal is presented in Figure 6(c). The obtained crystals are air sensitive as is expected for $Eu_3In_2P_4$, as previously such crystals were grown using flux method with excess In [46].

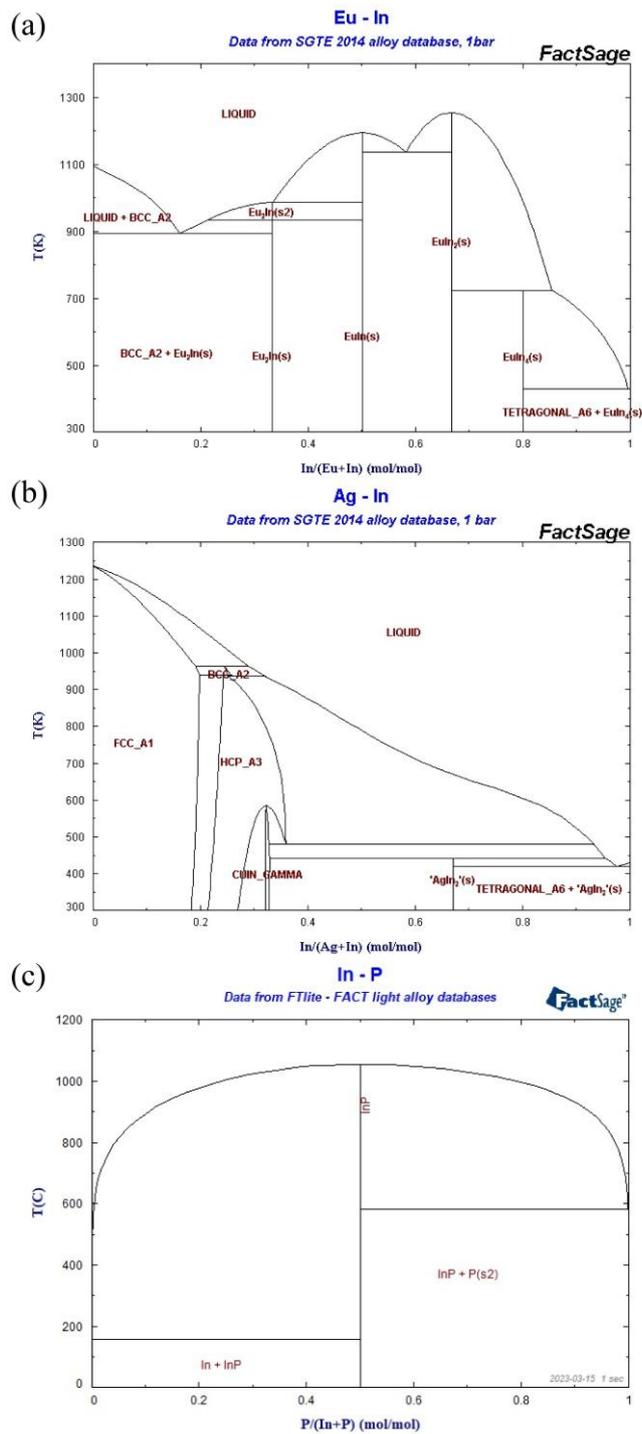

*Figure 11. Binary phase diagrams: (a) Eu-In, (b) Ag-In, (c) P-In* [40]

# 5 Conclusions

Flux methods are very convenient synthesis methods to grow single crystals, however, it is often not simple to predict whether crystals of the desired stoichiometry will be obtained at the very end. There are many factors that influence the result, such as the right choice of flux type, its amount, stoichiometry of ingredients, crucible material, heating profile, and the amount of compound to be synthesized. Furthermore, growing crystals of compounds that appear to be very similar, e.g., trying to create an analogy between arsenides and phosphides, is not straightforward.

We were able to reproduce the synthesis of EuAgAs from Bi flux. However, despite several attempts, synthesizing single crystals of EuAgP has not yet been successful. Nevertheless, the study demonstrates how the choice of fluxes such as: Bi, Sn, Pb, and In, and their chosen ratio influence the final reaction result.

# Acknowledgments

This work was financial supported by National Science Centre, Poland (Grant No. 2018/30/E/ST3/00377). W.T. acknowledges financial support National Science Centre, Poland (Grant No. OPUS: 2021/41/B/ST3/03454). The research project was partly supported by the program "Excellence initiative–research university" (projects 9907 and 6387) for the AGH University of Krakow.

# Author contributions

**Karolina Podgórska:** Formal analysis, Investigation, Writing – original draft preparation. **Damian Rybicki:** Conceptualization, Supervision, Writing – review & editing. **Lan Maria Tran:** Writing – review & editing. **Wojciech Tabiś:** Supervision, Writing – review & editing. **Łukasz Gondek:** Investigation, Formal analysis. **Michał Babij:** Conceptualization, Investigation, Supervision, Writing – review & editing.

# Funding sources

This work was financial supported by National Science Centre, Poland (Grant No. 2018/30/E/ST3/00377). W.T. acknowledges financial support National Science Centre, Poland (Grant No. OPUS: 2021/41/B/ST3/03454). The research project was partly supported by the program "Excellence initiative–research university" (projects 9907 and 6387) for the AGH University of Krakow.